\documentclass[]{jfm}

\usepackage{amsmath}
\usepackage{amssymb}
\usepackage{nicefrac}
\usepackage{bm}
\usepackage{graphicx}
\usepackage{import}
\usepackage{float}
\usepackage{tabularx}
\usepackage{pgf,tikz}
\usepackage{pgfplots}
\usepackage{gensymb}

\graphicspath{{Figures/}}

\begin{document}
\renewcommand{\vec}[1]{\bm{#1}}
\definecolor{bluegray}{cmyk}{0.25,0.25,0.16,0.0}

\newtheorem{lemma}{Lemma}
\newtheorem{corollary}{Corollary}

\shorttitle{Locally linear embedding} 
\shortauthor{A.~Ehlert and friends} 

\title{Locally linear embedding for transient cylinder wakes}

\author{Arthur Ehlert\aff{1}\footnote{Now at: Industrial Analytics IA GmbH, 10178 Berlin, Germany},
        Christian N. ~Nayeri\aff{1}, 
        Marek Morzynski\aff{2}, and 
        Bernd R.~Noack\aff{3,1}\corresp{\email{Bernd.Noack@limsi.fr}},}

\affiliation
{
\aff{1}
Institut f\"ur Str\"omungsmechanik und Technische Akustik (ISTA),
Technische Universit\"at Berlin,
M\"uller-Breslau-Stra{\ss}e 8,
10623 Berlin, Germany
\aff{2}
Pozna\'n University of Technology,
Chair of Virtual Engineering, 
Jana Pawla II 24, PL 60-965 Pozna\'n, Poland
\aff{3}
LIMSI,
CNRS, Universit\'e Paris-Saclay, B{\^a}t 507, rue du Belv\'ed\`ere, Campus Universitaire,
F-91403 Orsay, France
}

\maketitle \sf
   \begin{abstract} 
Reduced-order representations 
of an ensemble of cylinder wake transients are investigated.
Locally linear embedding identifies a two-dimensional manifold 
with a maximum error of 1\% from new snapshot data.
This representation outperforms 
a 50-dimensional POD expansion from the same data
and is not obtainable with cluster-based coarse graining of similar order.
This manifold resolves the steady solution, 
the stability eigenmodes, 
the first post-transient POD modes,
the intermediate vortex shedding structure as well as higher harmonics.
The snapshot data are generated by a direct numerical simulation 
of the two-dimensional wake behind a circular cylinder at a Reynolds number of 100. 
The initial conditions of the 16 transients start 
near the steady solution and converge to the period vortex shedding.
Many oscillatory flows can be expected to be characterized 
on two- or low-dimensional manifolds identifiable with locally linear embedding.
These manifolds have unexplored potential for prediction, estimation and control.
\end{abstract}
 
\sf
\section{Introduction}
\label{Sec:Introduction}
We use Locally Linear Embedding (LLE) to construct 
a two-dimensional manifold 
of the transient cylinder wake with negligible error. 
Reduced-order representations are a core goal of theoretical fluid mechanics.
Reduced-order models allow 
for a crisp understanding of coherent structure dynamics, 
for significant compression of flow data,
for computational inexpensive exploration of dynamics and
for model-based or model-inspired control. 
For low-dimensional dynamics and control, 
the construction of a least-order state space is pivotal.
Every degree of freedom can act as noise amplifier 
making the model less robust.
The difference between two-dimensional and three-dimensional models,
for instance,
is the difference between regular and potentially chaotic solutions.

We choose the two-dimensional cylinder wake as simple, 
yet surprisingly challenging benchmark example.
Over one hundred years ago,
\citet{Foeppl1913} approximated the steady vortex bubble 
with a vortex model.
Two years before,
\citet{Karman1911} explained the periodic shedding
with his celebrated vortex model 
associating the vortex street with his name.
In subsequent decades both reduced-order models
have experienced numerous refinements \citep{Lin1954proc,Timme1959rep},
have inspired alternative models or applications \citep{Faxen1927},
or have been used as control plant \citep{Protas2004pf}.
With the pioneering POD model for wall turbulence by \citet{Aubry1988jfm},
data-driven low-dimensional POD Galerkin models were also developed
for the oscillatory  two-dimensional \citep{Deane1991pfa} 
and three-dimensional cylinder wakes \citep{Ma2002jfm}.

The transient cylinder wake came into focus 
as a candidate for Landau's celebrated low-dimensional model 
for supercritical Hopf bifurcations \citep{Landau1944}.
It was left to \citet{Stuart1958jfm} to explain
the cubic damping term as interaction between fluctuation 
and distorted mean flow.
Decades later, the Stuart-Landau model was corroborated 
by the first global analyses of the steady cylinder wake solutions 
\citep{Zebib1987jem,Jackson1987jfm},
by experiments \citep{Schumm1994jfm}, and 
by a data-driven mean-field Galerkin model \citep{Noack2003jfm}.
None of the weakly nonlinear stability theories 
anticipated the dramatic impact of mode deformation on the dynamics
\citep{Zielinska1995pf,Siegel2008jfm,Loiseau2018jfm}.
These efforts strongly corroborate 
that the transient flow does not lie in a two-dimensional plane
spanned by two fixed spacial modes, as initially assumed by Landau, 
but are described by a two-dimensional Grassmann manifold.
Such manifolds have hitherto been hand-crafted 
from data and theoretical understanding.

Intriguingly,  proper orthogonal decomposition 
requires about 50 modes to resolve the transient 
at $Re=100$ with 1\% energy error or about 10\% amplitude error \citep{Loiseau2018jfm}.
Myriad of other modal expansions have been developed,
like Dynamic Mode Decomposition (DMD) \citep{Rowley2009jfm,Schmid2010jfm},
recursive DMD \citep{Noack2016jfm}, and variants thereof,
but all  global Galerkin expansions are challenged
by the deformation of modes with changed short-term averaged flow.
Only state-dependent modes \citep{Morzynski2006aiaa,Siegel2008jfm}
allow to resolve the flow on a two-dimensional manifold.

In this study, 
we formulate the manifold construction
quite generally as unsupervised autoencoding from snapshot data .
Autoencoding comprises a rich spectrum 
for low-dimensional parameterizations of high-dimensional data.
Examples range from proper orthogonal decomposition on linear subspaces
to nonlinear encoding and decoding with neural networks \citep{Ng2011proc}. 
A very popular technique for manifold learning
is locally linear embedding (LLE) by \citet{Roweis2000s}.
LLE has been extensively applied in machine learning, 
e.g. facial identification, but is rarely found in fluid mechanics.

In this study, we apply LLE
for the first accurate and purely data-driven manifold representation from snapshot data.
In addition, we  augment LLE to an autoencoder for manifold learning
with K-nearest neighbors as decoding method.
The manuscript is organized as follows.
Section \ref{Sec:Plant} describes the employed simulation data.
Locally linear embedding (LLE) and related data-driven reduced-order representations
are outlined in Section \ref{Sec:Method}.
In Section \ref{Sec:Results}, the transient cylinder wake data is 
encoded in a two-dimensional manifold using LLE.
Section \ref{Sec:Discussion} assesses the accuracy of LLE for validation data
and compares LLE with proper-orthogonal decomposition and clustering.
The last section \ref{Sec:Conclusions} summarizes the results
and concludes with an outlook for future research.

\section{Cylinder wake simulation}
\label{Sec:Plant}
The transient cylinder wake has been a surprisingly 
challenging benchmark of reduced-order representations
for over hundred years from vortex representations \citep{Karman1911}
via the POD Galerkin method \citep{Deane1991pfa}
to data-driven manifold models \citep{Loiseau2018jfm}.
In this section,  
the associated flow data from a  Direct Numerical Simulation (DNS) is described.
First, the configuration and simulation is presented in section \ref{Sec:DNS}.
In section \ref{Sec:Transient},
the simulation is illustrated.
An analysis of the aerodynamic force  in section \ref{Sec:ForceCoefficients}
suggests that the flow data may lie on a two-dimensional manifold,
consistent with the feature-based manifold by \citet{Loiseau2018jfm}.

\subsection{Direct numerical simulation}
\label{Sec:DNS}
The two-dimensional viscous, incompressible wake
behind a circular cylinder is computed.
This flow is characterized  by the Reynolds number
$Re = UD/\nu$ where $D$ represents the cylinder diameter,
$U$ the oncoming velocity, and 
$\nu$ the kinematic viscosity of the fluid.
The reference Reynolds number is set to $Re=100$,
which is significantly above the onset of vortex shedding at $Re=47$ \citep{Zebib1987jem,Jackson1987jfm}
and also far below the onset of three-dimensional instabilities around $Re =160$ \citep{Zhang1995pf,Barkley1996jfm}.

In the following, 
all quantities are assumed to be normalized
with the cylinder diameter $D$, the oncoming velocity $U$ and the density of the fluid $\rho$.
The two-dimensional cylinder wake is described 
by a Cartesian coordinate system $(x,y)$
with the origin in the cylinder center,
the $x$-axis pointing in streamwise
and the $y$-axis in transverse direction.
The incompressibility condition and Navier-Stokes equations read
\begin{subequations}
\begin{eqnarray}
\label{Eqn:Incomressibility}
\nabla \cdot \bm{u} &=& 0, \\
\partial_t \bm{u} + \bm{u} \cdot \nabla \bm{u} &=& - \nabla p + \frac{1}{Re} \triangle \bm{u},
\label{Eqn:NSE}
\end{eqnarray}
\end{subequations}
where $p$ represents the pressure, `$\partial_t$' partial differentiation with respect to time,
`$\nabla$' the Nabla operator and `$\cdot$' an inner product or contraction in tensor algebra.

The rectangular computational domain $\Omega_{\rm DNS}$
has a length and width of 50 and 20 diameters, respectively.
The cylinder center has a distance of 10 diameter to the front and lateral sides.
Summarizing,
$$ \Omega_{\rm DNS} = \left\{ 
   (x,y) \in {\cal R}^2
   \colon x^2+y^2 \le 1/4 
   \wedge -10 \le x \le 40 
   \wedge  \vert y \ \vert \le 10 \right\}.
$$
On the cylinder, the no-slip condition $\bm{u}=\bm{0}$  is enforced.
At the front $x=-10$ and lateral sides of the domain $y = \pm 10$ , 
a uniform oncoming flow $\bm{u}_{\infty} = (1,0)$ is assumed.
A vanishing stress condition is employed at the outflow boundary $x=40$.

\begin{figure}
	\centering
	\def\svgwidth{0.65\linewidth}
	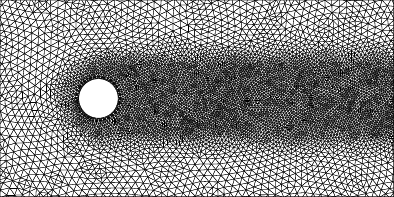
	\caption{Close-up view of the grid around the cylinder.}
	\label{Fig:GridCloseUp}
	\centering
\end{figure}

Simulations are performed with a finite-element method
on an unstructured grid with implicit time integration.
This solver is third-order accurate in time and and second-order accurate in space.
Details about the Navier-Stokes and stability solvers 
are described in \citet{Morzynski1999cmame,Noack2003jfm}.
The triangular mesh consists of $59112$ elements. 
Figure \ref{Fig:GridCloseUp} shows a close-up view of the grid around the cylinder.

The employed initial conditions 
are based on the unstable steady solution $\bm{u}_s$
and a small disturbance with the most unstable eigenmode $\bm{f}$.
The steady solution is computed with a Newton gradient solver.
The eigenmode computation is described in our earlier work \citep{Noack2003jfm}.
The disturbance is the real part of the product of the eigenmode and unit phase factor $e^{\imath \phi}$.
Here, `$\imath$' denotes the imaginary unit and $\phi$ the phase. 
The amplitude $\epsilon$ is chosen to create a perturbation 
with a fluctuation energy of $10^{-4}$.
The resulting initial condition reads
\begin{equation}
 \bm{u}(\bm{x},t=0) = \bm{u}_s ( \bm{x} ) 
   + \epsilon \Re \left\{ \bm{f}_1 ( \bm{x} ) \> e^{\imath \phi} \right\} .
\label{Eqn:InitialCondition}
\end{equation}
16 initial conditions are considered.
These correspond to equidistantly sampled  phases 
 $\Phi\in[22.5\degree,45\degree,\dots,337.5\degree,360\degree]$.
Integration is performed from $t=0$ to $t=200$ capturing the complete transient and post-transient state.
The time step is $\Delta t = 0.1$ corresponding to roughly one 50th of the period.

\subsection{From the steady solution to limit-cycle dynamics}
\label{Sec:Transient}

In this section, 
the transients from the steady solution
to periodic vortex shedding are investigated.
The flow is analyzed in the observation domain 
\begin{equation}
\Omega :=  \left\{ (x,y) \in \Omega_{\rm DNS} :  5 \le x\le 15 \wedge 5 \le y \le 5 \right\} .
\label{Eqn:ObservationDomain}
\end{equation}
This domain is about twice as long as the vortex bubble of the steady solution.
The streamwise extent is large enough to resolve over one wavelength of the initial vortex shedding 
as characterized by the stability eigenmode.
A larger domain is not desireable, 
because a small increase in wavenumber 
during the transient will give rise to large phase differences in the outflow region,
complicating the comparison between flow states.
The domain is consistent with earlier investigations 
by the authors \citep{Noack2003jfm,Gerhard2003aiaa}
and similar to the domain of other studies \citep{Deane1991pfa}.

The analysis is based on the inner product
of the Hilbert space of square-integrable functions
over the observation domain $\Omega$.
This inner product between two velocity fields $\bm{v}$
and $\bm{w}$ is defined by 
\begin{equation}
\left( \bm{v},\bm{w} \right)_{\Omega}
= \int\limits_{\Omega} \!\! d\bm{x} \> \bm{v} \cdot \bm{w} 
\label{Eqn:InnerProduct} 
\end{equation}
where `$\cdot$' denotes the Euclidean inner product.
The corresponding norm of the velocity field $\bm{v}$ reads
\begin{equation}
\Vert \bm{v} \Vert_{\Omega} 
= \sqrt{\left( \bm{v}, \bm{v}  \right)_{\Omega}}.
\label{Eqn:Norm} 
\end{equation}

The flow $\bm{u}$ is decomposed 
into a slowly varying base flow $\bm{u}^B$
and an oscillatory fluctuation $\bm{u}^\prime$,
\begin{equation}
\bm{u} = \bm{u}^B + \bm{u}^\prime.
\label{Eqn:Decomposition}
\end{equation}
The base flow is defined as the projection of the flow
on the line connecting  the steady solution $\bm{u}_s$ 
and the post-transient mean flow $\bm{u}_0$.
In other words,
\begin{equation}
\bm{u}^B(\bm{x},t) = \bm{u}_s (\bm{x}) 
 + a_{\Delta} (t) \> \bm{u}_{\Delta} (\bm{x} ), 
\label{Eqn:BaseFlow}
\end{equation}
with the shift-mode 
$\bm{u}_{\Delta} = \left ( \bm{u}_0 - \bm{u}_s \right) / 
\Vert  \bm{u}_0 - \bm{u}_s \Vert_{\Omega}$
and amplitude $a_{\Delta} = \left ( \bm{u}-\bm{u}_s,  \bm{u}_{\Delta} \right )_{\Omega}$.
This definition approximates a short-term averaged flow
and generalizes the notion in the stability literature 
where the steady solution is identified with the base flow.

The shift-mode amplitude $a_{\Delta}$ characterizes
the mean-flow distortion \citep{Stuart1958jfm}
while the fluctuation energy 
\begin{equation}
\mathcal{K} :=  \left\Vert \bm{u}^\prime \right \Vert_{\Omega}^2/2
\label{Eqn:TurbulentKineticEnergy}
\end{equation}
parameterizes the fluctuation level.
We also refer to $\mathcal{K}$ as \emph{turbulent kinetic energy} (TKE)
following the mathematical definition of statistical fluid mechanics,
realizing that the flow is laminar, not turbulent.

\begin{figure}
	\centering
	\input{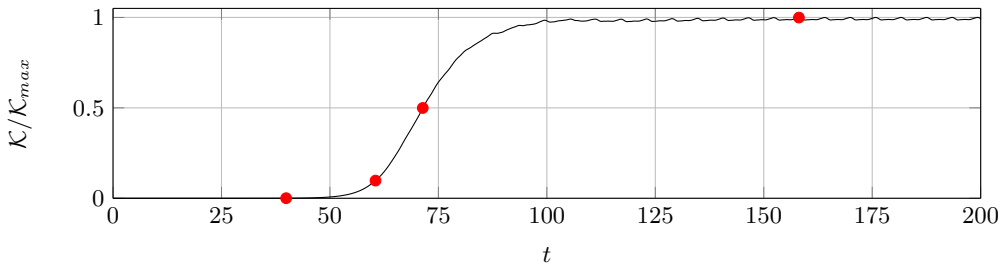}
	\caption{Evolution of the turbulent kinetic energy $\mathcal{K}$ with  time $t$ associated 
                 with an initial condition for $\phi=22.5 \degree$.
	The values are normalized with the maximum value $\mathcal{K}_{max}$.
	Red points indicate normalized fluctuation levels of $0$, $10$, $50$ and $100$ percent.} 
	\label{Fig:TKE}
	\centering
\end{figure}
Figure \ref{Fig:TKE} displays the TKE evolution with time.
The maximum TKE value $\mathcal{K}_{max}$ 
is used for normalization. 
Three dynamic phases can be distinguished. 
Within the first 30 convective time units the flow 
exhibits \emph{linear dynamics} 
or exponential growth in the neighbourhood of the steady solution.
This exponential growth can clearly 
be seen in a logarithmic plot \citep{Noack2003jfm}.
In the second, \emph{nonlinear transient phase} 
for  $50<t<100$  the flow transitions 
from the steady solution to the limit cycle 
with decreasing growth-rate.
In the \emph{post-transient phase} for $t>150$,
a periodic vortex shedding or, equivalently, limit-cycle dynamics is observed.
The figure marks four times for TKE levels 
near 0\%, 10\% , 50\% and 100\%,
corresponding to the linear dynamics phase,
the beginning and middle of the nonlinear transient phase 
and  the limit cycle.

\begin{figure}
	\centering
	\def\svgwidth{0.8\linewidth}
	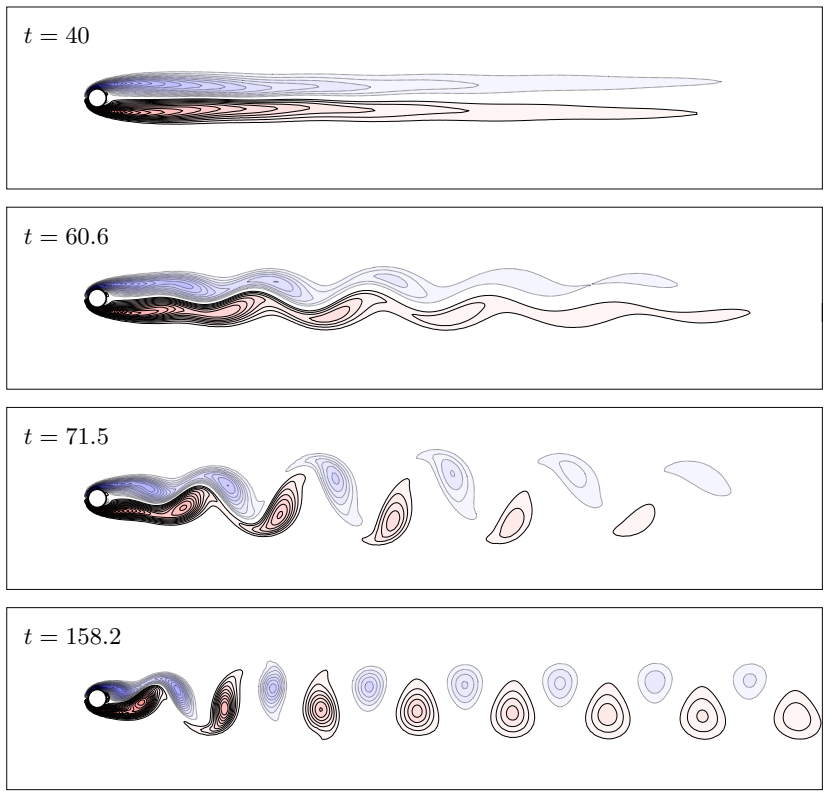 
\caption{Vorticity snapshots 
corresponding to $0$, $10$, $50$ and $100$ percent fluctuation level
for the simulation displayed in figure \ref{Fig:TKE}.
The flow is visualized by iso-contours of vorticity
with positive (negative) values marked by solid (dashed) lines and red (blue) background.
The iso-contour levels and color scales are the same for all snapshots.}
	\label{Fig:Snapshots}
\end{figure}

In figure \ref{Fig:Snapshots} the vorticity 
for the four selected time instants is shown.
Positive (negative) values of vorticity are shown in red (blue) 
bounded by solid (dashed) lines.
The three dynamic phases can be distinguished 
based on the closeness of vortex shedding to the cylinder
and on the formation of pronounced individual vortices.

This discussion provides a basis for the time interval
 $[t_{\rm min}, t_{\rm max}]$ for snapshot selection.
A lower bound $t_{\rm min}= 40$ is chosen.
This bound guarantees a TKE below $0.01$ \% or, equivalently $10^{-4}$ of the asymptotic maximum value. 
The upper bound $t_{\rm max}=110$ includes few periods on the limit cycle.

\subsection{Force coefficients}
\label{Sec:ForceCoefficients}
In this section,  
the drag and lift force coefficients are investigated. 
Drag $F_D$ and  lift $F_L$  per unit spanwise length are computed 
from pressure and skin-friction contributions on the cylinder.
The lift and drag coefficients were computed using  
\begin{equation}
C_i=\frac{F_i}{\frac{1}{2} \rho U_{\infty}  D}\mathrm{,}
\label{Eqn:LiftDrag}
\end{equation}
where $i$ represents `$D$' for drag and `$L$' for lift.

\begin{figure}
	\centering
	\input{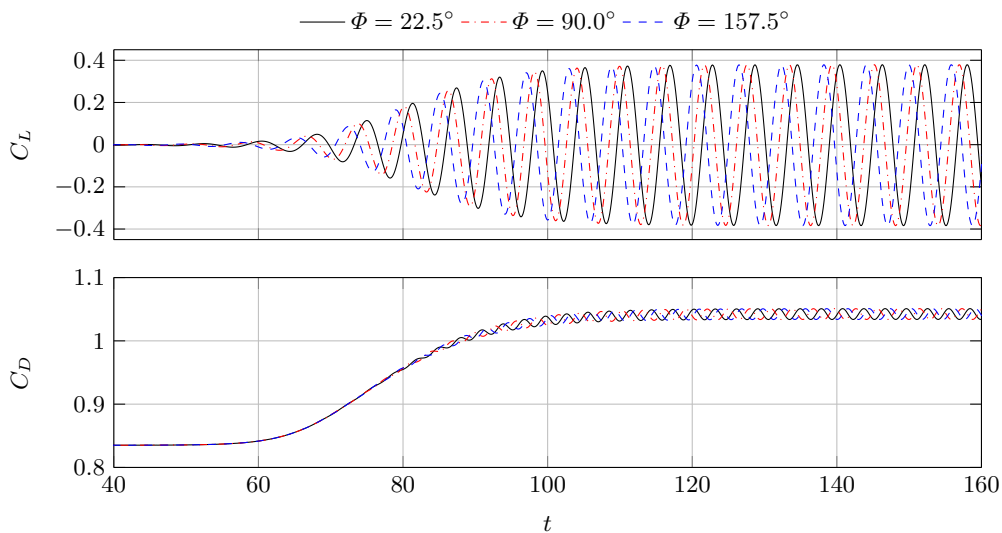}
	\caption{Evolution of the lift $C_L$ cand drag $C_D$ coefficients for three initial conditions
                 associated with the phases $\phi = 22.5 \degree$ of figures \ref{Fig:TKE} and \ref{Fig:Snapshots} (solid line),
                 $\phi = 90 \degree$ (dash-dotted line) and $\phi=157.5 \degree$ (dashed line).} 
	\label{Fig:LiftDrag}
\end{figure}
Figure \ref{Fig:LiftDrag} illustrates  the force coefficients
for three different initial conditions.
The phase $\phi$ of the initial condition evidently effects 
the phase of the lift coefficient,
while a single envelope bounds all fluctuations.
The phase shift in perturbation results in an equal phase shift of the evolution of the simulation.
The drag coefficient has a slow relaxational dynamics with a small fluctuation of the second harmonics.

\begin{figure}
	\centering
	\def\svgwidth{0.5\linewidth}
	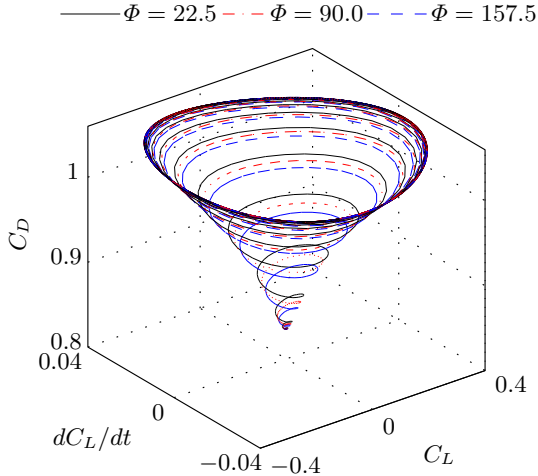
\caption{Phase portrait of the simulations shown in figure \ref{Fig:LiftDrag}.
The feature space is spanned by the drag coefficient, $C_D$, the lift coefficient $C_L$ and its derivative $dC_L/dt$.
The trajectories seem to be embedded in a conical two-dimensional manifold like in \citet{Loiseau2018jfm}.}
	\label{Fig:ForcesManifold}
\end{figure}

This oscillatory dynamics motivate a phase space
with the drag coefficient, lift coefficient and time-derivative of the latter. 
Following \citet{Loiseau2018jfm}, 
a phase portrait of these quantities is depicted
in figure \ref{Fig:ForcesManifold}.
The trajectories lie on a conus-like manifold.
These results motivate the application 
of  locally linear embedding as manifold learning technique.

\section{Autoencoders---From POD to LLE}
\label{Sec:Method}
This section describes  \emph{Locally Linear Embedding} (LLE) 
contrasting it to other data-driven reduced-order representations.
Starting point is a rich set of  snapshot data from transient velocity fields.
Typically, reduced-order representations employ an autoencoder:
Given an ensemble of snapshot data, 
find an \emph{encoder} from this data to a low-dimensional feature space 
and a \emph{decoder} from the feature vector to the original state space 
such that the composition of both mappings approximates an identity operation for the given data.

First (\S~\ref{Sec:Autoencoder}), the idea of an autoencoder is recapitulated.
Proper Orthogonal Decomposition (POD) and clustering constitute examples 
and are outlined in \S~\ref{Sec:POD} and \S~\ref{Sec:Clustering}, respectively.
For visualization purposes, 
two-dimensional proximity maps represent an appealing encoding methodology 
which may be augmented by a decoder as reviewed in \S~\ref{Sec:ProximityMap}.
All these reduced-order representations may be challenged by data on curved low-dimensional manifolds.
This data structure is the niche application of  LLE 
which incorporates aspects of POD, clustering and proximity maps.
The theoretical foundation of LLE is outlined in \S~\ref{Sec:LLE} 
and followed by two geometrical examples in \S~\ref{Sec:LLEExample}.

\subsection{Autoencoder of snapshot data}
\label{Sec:Autoencoder}
First, the very idea of an autoencoder is elaborated.
This idea is shared by POD, clustering, vortex representations 
and many other data-driven reduced-order representations.
Starting point is an ensemble of $M$ flow snapshots $\bm{u}^m (\bm{x})$, $m=1,\ldots,M$,
geometrically covering the area of interest.

An autoencoder targets a low-dimensional parameterization of the snapshot data, say in $\mathcal{R}^N$.
More precisely, an autoencoder comprises an encoder $G$ 
from the high- or infinite-dimensional state space to a low-dimensional feature space, 
e.g.\
\begin{equation}
\bm{u}^m \mapsto \bm{a}^m :=G \left (\bm{u}^m \right) \in \mathcal{R}^N, \quad m=1,\ldots, M 
\label{Eqn:Encoder}
\end{equation}
and a decoder or state estimator $H$, e.g.\ 
\begin{equation}
\bm{a}^m \mapsto \hat{\bm{u}}^m :=H \left (\bm{a}^m \right),  \quad m=1,\ldots, M .
\label{Eqn:Encoder}
\end{equation}
Ideally, the autoencoder identifies the best possible pair of encoder $G$ and decoder $H$
which minimize the in-sample error of the estimator/decoder
\begin{equation}
E_{in} := \frac{1}{M} \sum\limits_{m=1}^M  \left \Vert \hat{\bm{u}}^m - \bm{u}^m \right \Vert_{\Omega}^2.
\label{Eqn:InSampleError}
\end{equation}

We remark that the in-sample error is used 
for an automated calibration of sufficiently simple encoders and decoders.
The in-sample error $E_{in}$ should not be minimized with overly complex encoders/decoders
at the expense of the out-of-sample error $E_{out}$ for new data.
This would be overfitting, a violation of Occam's razor.
The goal of any data-driven model is to minimize the out-of-sample error 
for new data as beautifully elaborated by \citet{AbuMostafa2012book}.

\subsection{POD---Autoencoding with an affine subspace}
\label{Sec:POD}
POD can be considered as optimal linear autoencoder 
onto an affine $N$-dimensional subspace. 
Let $\bm{u}_0$ be the average of the snapshot ensemble, $\bm{u}_i$, $i=1,\ldots,N$, be the $N$ POD modes, and $a_i$ be the corresponding mode coefficients.
Then the encoder $G$ of a velocity field $\bm{u}$ 
to the mode amplitudes $\bm{a}= ( a_1, \ldots, a_N)^T $ (`$T$' denoting the transpose)
is defined by 
\begin{equation}
a_i := \left ( \bm{u}-\bm{u}_0, \bm{u}_i \right)_{\Omega}, \quad i=1,\ldots,N
\label{Eqn:PODCoefficient}
\end{equation}
while the decoder $H$ reads 
\begin{equation}
\hat{\bm{u}} (\bm{x}) =  \bm{u}_0 (\bm{x}) + \sum\limits_{i=1}^M a_i \bm{u}_i ( \bm{x} ).
\label{Eqn:PODExpansion}
\end{equation}
The optimality condition \citep[see, e.g.][]{Holmes2012book} implies 
a minimal in-sample error from equation \eqref{Eqn:InSampleError}.
We cannot find another autoencoder which yields a better $N$th order Galerkin expansion.
Evidently, many POD equivalent Galerkin expansions can be constructed by coordinate transformations.
Yet, POD also requires that the first $I \in \{1,\ldots,N\}$ modes optimally resolve the corresponding $I$-dimensional Galerkin expansions, i.e. the coordinates are sorted by relevance.
For generic snapshot data, the modes and coordinates are unique modulo a sign.

\subsection{Clustering---Autoencoding into bins of snapshots}
\label{Sec:Clustering}
The key idea of clustering is representing the snapshots by a small number, say $K$, of centroids $\bm{c}_k$ with $k=1,\ldots,K$.
Every snapshot $\bm{u}^m$ can be associated with its closest centroid $\bm{c}_k$.
Thus, the encoder $G$ maps the velocity field $\bm{u}$ to $k \in \{1,\ldots,K \}$, the index of the closest centroid. 
In other words, the encoder creates 'bins' of similar snapshots.
The decoder $H$ approximates the velocity field by the closest centroid $\hat{\bm{u}} = \bm{c}_k$.
The k-means algorithm aims to minimize the in-sample error \citep{Arthur2007proc}.
For generic data, the centroids can be expected to be unique modulo numbering.
Clustering with $K$ bins cannot yield a lower in-sample error 
than a POD representation with $K$ modes $\bm{u}_i$, $i=0,\ldots,K-1$.
Both clustering and POD span a $K-1$-dimensional subspace, 
but a POD expansion can interpolate states while the centroids are fixed.

\subsection{Proximity map---Cartographing the snapshots for visualization}
\label{Sec:ProximityMap}
The goal of a proximity map is to cartograph high-dimensional snapshots 
in a visually accessible, often two-dimensional feature space which preserves neighbourhood relations as well as possible.
Let 
$\bm{\gamma}^m =\left( \gamma_1^m, \gamma_2^m \right)^T \in \mathcal{R}^2$ 
with $m=1,\ldots, M$ be the two-dimensional feature vectors 
corresponding to the snapshots $\bm{u}^m, m=1,\ldots, M$.
In classical multidimensional scaling (CMDS) \citep{Cox2000book}, 
these features minimize the accumulative error of the distances between the snapshots
\begin{equation}
E = \sum\limits_{m=1}^M \sum\limits_{n=1}^M \left [ \left \Vert \bm{u}^m - \bm{u}^n \right \Vert - \left \Vert \bm{\gamma}^m - \bm{\gamma}^n \right \Vert\right ]^2.
\label{Eqn:AccumulativeErrorCMDS}
\end{equation}
The translational degree of freedom is removed by requesting centered features,
\begin{equation}
\sum\limits_{m=1}^M  \bm{\gamma}^m = 0.
\label{Eqn:CenteredFeatures}
\end{equation}
The rotational degree of freedom is fixed by requiring  
the first feature coordinate to be maximum.
In general, 
for an $N$-dimensional feature space, 
the sum of first $I$ variances is maximized for all $I \in \{1,\ldots,N \}$.
For the invariance of the error under mirroring, however, there is no cure, 
like with the sign indeterminacy of POD modes and amplitudes.
In fact, the resulting proximity map yields the first two POD amplitudes $a_1$, $a_2$.
The resulting metric may be tailored to specific applications, 
e.g. identifying regions with similar cost functions \citep{Kaiser2017tcfd}.
Since proximity maps are based on preserving neighbourhood information, 
it is strongly related to LLE. 

\subsection{LLE---Representing low-dimensional manifolds}
\label{Sec:LLE}
LLE \citep{Roweis2000s} targets a low-dimensional approximation of $M$ typically high-dimensional data points on a manifold preserving the local neighbourhood information as well as possible. 
In particular, neighbouring points in the original data space remain neighbours in the low-dimensional embedding space. 
The idea of the LLE algorithm  can be inferred from the following three steps:
\begin{enumerate}
\item Compute the $K$ nearest neighbours of each point.
Here, the distance between the snapshots is computed 
with the Hilbert-space norm from equation \eqref{Eqn:Norm}. 
Let $n^m_k$ with $k=1,\ldots,K$ 
be the indices of the snapshots closest to the $m$th one.
\item Find the weights $w_{mn}$ which optimally reconstruct each point from its $K$ neighbours.
\begin{equation}
\bm{u}^m  \approx \sum\limits_{n=1}^M w_{mn} \bm{u}^{n}
\label{Eqn:OptimalWeights}
\end{equation}
The non-negative weights $w_{mn}$ of more distant snapshots vanish, 
i.e. the right-hand side is a sum over the $K$ neighbours.
Moreover, the weights add up to unity.
Summarizing,
\begin{subequations}
\begin{eqnarray}
\label{Eqn:WeightSupport}
w_{mn} &\ge& 0, \\
\sum\limits_{m=1}^M w_{mn} &=& 1 \quad \rm{and}\\
w_{mn} &=& 0 \quad \hbox{if} \quad  n \not \in \left \{ n_1^m, \ldots, n_K^m \right\}  .
\end{eqnarray}
\end{subequations}
\item Compute low-dimensional coordinates $\bm{\gamma}^m \in \mathcal{R}^N$ 
to mimic the neighbourhood relationships as good as possible.
\begin{equation}
\bm{\gamma}^m  \approx \sum\limits_{n=1}^M w_{mn} \bm{\gamma}^n
\label{Eqn:OptimalEmbedding}
\end{equation}
\end{enumerate}

The weights from equations \eqref{Eqn:OptimalWeights} 
and \eqref{Eqn:OptimalEmbedding} minimise two cost functionals.
Firstly, the weight matrix $\bm{W} = \left ( w_{mn} \right ) \in \mathcal{R}^{M \times M}$ 
minimizes the \emph{residual sum of squares} (RSS)
\begin{equation}
\hbox{RSS} (\bm{w})=\sum\limits_{m=1}^M  
   \left \Vert \bm{u}^m - \sum\limits_{n=1}^M   w_{mn} \bm{u}^n \right \Vert_{\Omega}^2,
\label{Eqn:RSS}
\end{equation}
subject to the weight constraints \eqref{Eqn:WeightSupport}.
Additionally, the LLE coordinates minimize the \emph{reconstruction error} (RE)
\begin{equation}
\hbox{RE} (\bm{\Gamma})=\sum_{m=1}^M
        \Vert \bm{\gamma}^m - \sum\limits_{n=1}^M w_{mn}  \bm{\gamma}^n \Vert^2,
\label{Eqn:RE}
\end{equation}
satisfying a centering constraint
\begin{equation}
\sum\limits_{m=1}^M \bm{\gamma}^m =0
\label{Eqn:CenteringGamma}
\end{equation}
and the constraint of the covariance matrix being the identity matrix
\begin{equation}
\bm{\Gamma}^T\bm{\Gamma}=\bm{I} \in \mathcal{R}^{N \times N}.
\label{Eqn:CovarianceIdentity}
\end{equation}
The centering removes the translational degree of freedom.
The constraint of the co-variance matrix prevents the trivial solution $\bm{\gamma}^m \equiv \bm{0}$, 
enforces the same variation of all  coordinates and guarantees different coordinates to be uncorrelated.
This behaviour is desirable.
The resulting LLE coordinates $\bm{\gamma}^m \in \mathcal{R}^N$ are comprised in a $M\times N$ matrix.
\begin{equation}
\bm{\Gamma} :=
\left( \begin{array}{ccc} 
       \vert         & \cdots & \vert 
    \\ \bm{\gamma}^1 & \cdots & \bm{\gamma}^M
    \\ \vert         & \cdots & \vert 
       \end{array}
\right).
\end{equation}

\subsection{LLE---Introductory examples}
\label{Sec:LLEExample}
\begin{figure}
	\centering
	\def\svgwidth{0.75\linewidth}
	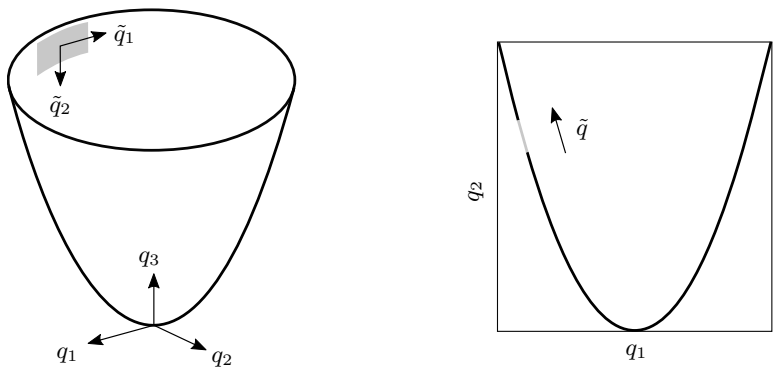
\caption[Two examples of manifolds.]
	{Paraboloid and parabola as illustration examples. 
         Left: Three-dimensional paraboloid with a patch that can be approximated using the two coordinates $\tilde{q}_1$ and $\tilde{q}_2$.
	Right: Two-dimensional parabola with a segment that can be approximated using  one coordinate $\tilde{q}$.}
	\label{Fig:Examples}
\end{figure}

In the following, 
LLE is illustrated with a part of a paraboloid  $q_3= q_1^2+q_2^2$, $q_3 \le 1$ 
and of a parabola from $q_2=q_1^2$, $q_2 \le 1$.
The data are randomly placed points on these objects.
In figure \ref{Fig:Examples} schematics of the three and two-dimensional shapes are depicted.
The examples illustrate the concept of a manifold. 
Locally, the paraboloid can be approximated 
by a two-dimensional coordinate system $\tilde{q}_1$ and $\tilde{q}_2$.
The parabola only requires one coordinate locally $\tilde{q}$.\\

\begin{figure}
	\centering
	\def\svgwidth{0.9\linewidth}
	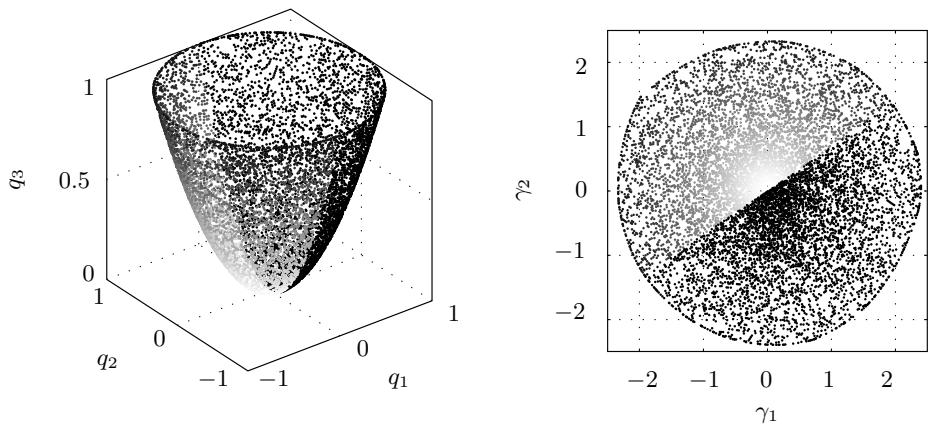
\caption{Locally linear embedding (LLE) of random points on a paraboloid. 
         Randomly sampled points $\bm{q} = (q_1,q_2,q_3)$ (left) 
         and the first two embedding coordinates $\bm{\gamma}=(\gamma_1,\gamma_2)$ 
         from LLE using ten nearest neighbours (right).
         The points and their LLE coordinates are color coded as described in the text.}
	\label{Fig:Paraboloid}
\end{figure}

\begin{figure}
	\centering
	\def\svgwidth{0.65\linewidth}
	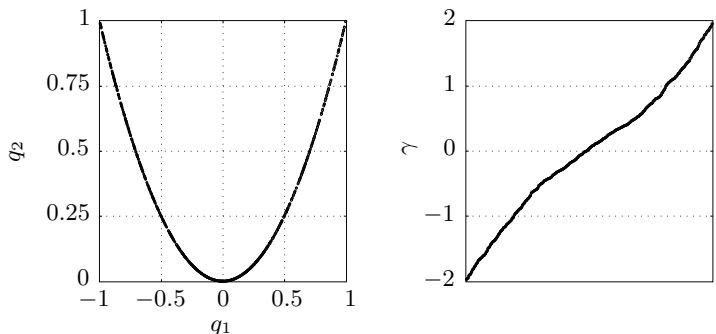
\caption{LLE of random points on a parabola.
         Randomly sampled points $\bm{q}=(q_1,q_2)$  (left) and the first embedding coordinate $\gamma$  
         from LLE using ten nearest neighbours (right).
         The abscissa of the right figure corresponds to the index of the sorted LLE coordinate.}
	\label{Fig:Parabola}
\end{figure}

$M=9,000$ points $\bm{q}^m$ are randomly sampled on the paraboloid 
resulting in an approximately uniform point density in the $q_1$--$q_2$ plane.
The parabola is sampled with $M=750$ random points, 
analogously with near-uniform distribution along the $q_1$ axis.
A color coding has been applied to the sampled points in order to reveal the functionality of LLE.
The points on the parabola are color-coded according to $q_1$ (figure \ref{Fig:Parabola}).
The paraboloid (figure \ref{Fig:Paraboloid}) is split into two halves: 
for $q_1>0$ all points are black whereas the other half is defined by a greyscale according to $q_3$.
For the paraboloid, 
LLE maps the sampled points onto a circular region (figure \ref{Fig:Paraboloid}, right).
The LLE coordinates $\gamma_1$, $\gamma_2$ roughly represent a rotated 
and re-scaled version of the first two original coordinates $q_1$, $q_2$.
The radial coordinate is nearly proportional to the arc length of the geodesic 
from the origin $\bm{0}$ to the point $\bm{u}$ on the paraboloid. 
Hence, the upper region of the paraboloid occupies---relative to the total projection area $q_1^2+q_2^2 \le 1$---a larger area 
in the $\bm{\gamma}$-plane as compared to its projection 
onto the $q_1$-$q_2$ plane. 
For the parabola on the other hand (figure \ref{Fig:Parabola}), 
the LLE coordinate roughly scales with the arc length.

LLE has identified the encoder 
and a $K$-nearest neighbour regression could be used for a decoder, 
from arbitrary coordinates in the data range to the higher-dimensional data space.
Both for the paraboloid and the parabola, 
the existence of such decoders is frequently observed for simple examples 
but it is not guaranteed per se.
One counterexample is a one-dimensional LLE for a circle in a plane.

LLE has notable similarities and differences with POD, clustering and the proximity map.
All encoders are based on snapshot data 
which is expected to cover a region of interest in the state space.
The snapshots should be reasonably dense 
but are not expected to be time-resolved or sorted unlike DMD \citep{Schmid2010jfm,Rowley2009jfm}.
POD and clustering come with decoders by definitions.
LLE and proximity maps may easily be upgraded with a decoder.
A simple decoder uses the $K$-nearest neighbours (KNN)
of a feature vector of existing data $\bm{\gamma}^{n_k}$, $k=1,\ldots,K$ 
and  applies the same weights to construct the state $\bm{u}^{n_k}$, $k=1,\ldots,K$ \citep[see,e .g.][]{Loiseau2018jfm}.
There is no guarantee that satisfying encoders exist,
while POD and clustering can offer error estimates 
of their low-dimensional flow representation.
From a geometric point of view, POD is a natural choice for linear subspaces, 
clustering for data with multi-modal distribution and LLE for manifolds. 
In our study, we will show the distinct advantage of LLE applied 
to oscillatory transients to a stable limit cycle.
 
\section{LLE of cylinder wake transients}
\label{Sec:Results}

We  present the application of locally linear embedding (LLE)
to the discussed data of the cylinder wake transients. 
First (\S~\ref{Sec:Results:Data}), the snapshot data is defined.
In \S~\ref{Sec:Results:LLE} this data is shown in the feature space.
Section \ref{Sec:Results:Force} provides an interpretation 
of LLE coordinates in terms of the force coefficients.
In \S~\ref{Sec:Results:Autencoder}, an inverse mapping (decoder)
from the  feature vector to the velocity field is introduced.
The corresponding LLE-based autoencoder yields 
an out-sample-error of less than one percent.

\subsection{Snapshot data}
\label{Sec:Results:Data}
Following section \ref{Sec:Plant}, 
we employ snapshots from all 16 transients.
Spatially, the velocity field is  constrained to the observation domain \eqref{Eqn:ObservationDomain}.
Temporally, only snapshots from  time interval $[t_{\rm min}, t_{\rm max}] = [40,160]$ are taken.
The time step between two consecutive snapshots is $\Delta t = 0.1$.
This gives rise to a total of $16 \times 1200 =  19,200$ snapshots.

\subsection{Two-dimensional embedding using LLE}
\label{Sec:Results:LLE}

\begin{figure}
	\centering
	\def\svgwidth{0.45\linewidth}
	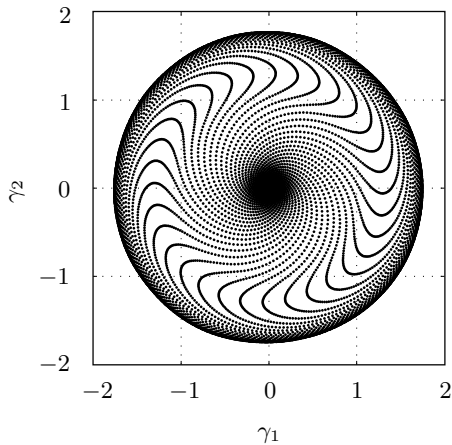
\caption{LLE of 16 cylinder wake transients. 
         The figures displays the first two embedding coordinates $\bm{\gamma}=(\gamma_1,\gamma_2)$ 
          resulting from $K=15$ nearest neighbours.}
	\label{Fig:lleK15}
	\centering
\end{figure}
\begin{figure}
	\centering
	\def\svgwidth{0.8\linewidth}
	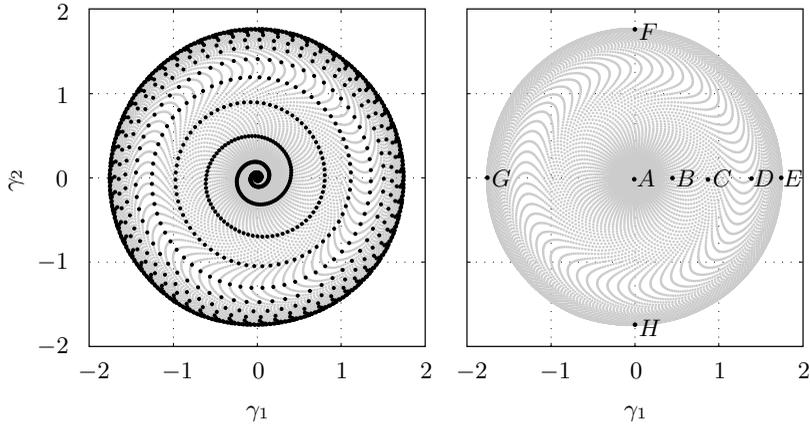
\caption{Same as figure \ref{Fig:lleK15}, but with a single trajectory (left) and few features points (right) highlighted.
         The grey points correspond to the coordinates shown in figure \ref{Fig:lleK15}.
         The trajectory is emphasized with black points 
         and the points $A$--$H$ are marked for later inspection.}
	\label{Fig:lleK15Marked}
\end{figure}
\begin{figure}
	\centering
	\def\svgwidth{0.8\linewidth}
	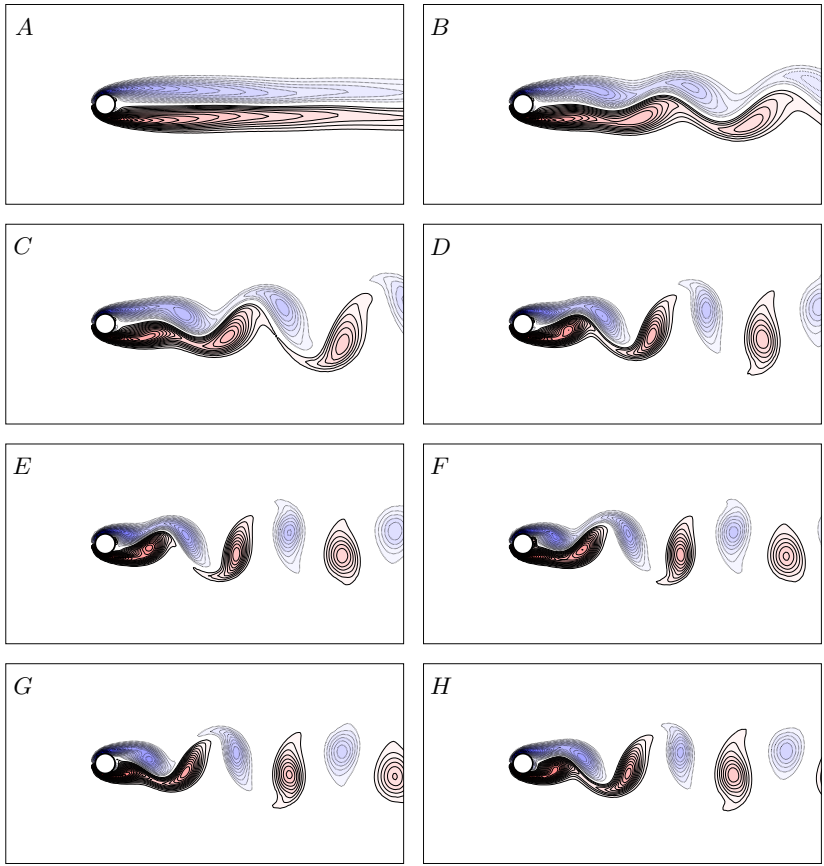
\caption{Vorticity snapshots corresponding to marked points in figure \ref{Fig:lleK15Marked}. 
The flow is visualized like in figure \ref{Fig:Snapshots} with the same iso-contour values and color code.
From a comparison with figure \ref{Fig:lleK15Marked},
the amplitude $\Vert \bm{\gamma} \Vert$ correlates with the fluctuation energy
while the phase $\hbox{atan} (\gamma_2/\gamma_1)$ corresponds to the shedding phase.}
	\label{Fig:vortMarked}
	\centering
\end{figure}
Figure \ref{Fig:lleK15} displays the  two-dimensional LLE embedding using $K=15$ nearest neighbours. 
Each point corresponds to one snapshot of the 16 transients.
The trajectories $t \mapsto \bm{\gamma}$ start near the origin 
and spiral outwards to a circular limit cycle with radius $1.75$.
A single trajectory is highlighted in  figure \ref{Fig:lleK15Marked} (left)
showing about eight revolutions before reaching the limit cycle.
Eight data points $A$--$H$ are marked in right side of figure \ref{Fig:lleK15Marked}.
The first four points $A$--$D$ correspond feature amplitudes of $0$, $25$, $50$ and $75$\%,
the next four points $E$--$H$ have angles $0$, $90^\circ$, $180^\circ$ and $270^\circ$.
Thus, the radial and angular dependencies can be explored.
The corresponding velocity fields are visualized  in figure \ref{Fig:vortMarked}.
The origin $A$, corresponds to the steady solution.
Points $B$, $C$ and $D$ illustrate increasing levels of fluctuations;
the vortex shedding moves upstream towards the cylinder.
Points $E$--$H$ correspond to different phases of vortex shedding 
at quarter-period increments.
Locally linear embedding provides an astonishingly clear 
two-dimensional representation of the cylinder wake transients.
The feature vector clearly resolves 
amplitude and phase purely from the snapshot data
and without imposing any prior knowledge of an oscillatory process.

\subsection{Force coefficients in LLE}
\label{Sec:Results:Force}
\begin{figure}
	\centering
	\def\svgwidth{0.80\linewidth}
	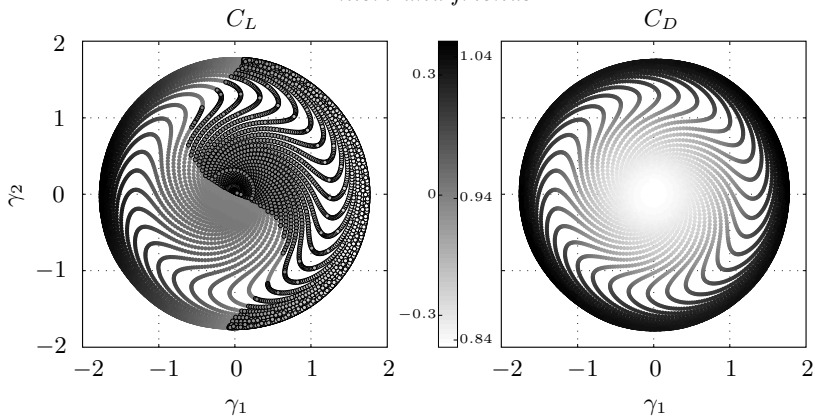
\caption{Same as figure \ref{Fig:lleK15}, but color-coded with the drag $C_D$ and lift coefficient $C_L$.
The grey scale in the middle denotes the value.
Negative values of the lift coefficients are emphasized by black contours of the corresponding circles.}
	\label{Fig:lleForces}
	\centering
\end{figure}
Figure  \ref{Fig:lleForces} displays how the lift and drag forces are related to the feature space.
The lift and drag coefficients are color-coded according the bar in the middle.
Negative values of the lift are darkened by a black contour of the bullet.
The lift strongly correlates with $\gamma_1$ displaying 
the minimum (maximum) as rightmost (leftmost) point on the limit cycle.
Drag increases with the feature amplitude,
assuming its minimum at the origin and maximum on the limit cycle.

\subsection{LLE-based autoencoder}
\label{Sec:Results:Autencoder}
\begin{figure}
	\centering
	\def\svgwidth{0.9\linewidth}
	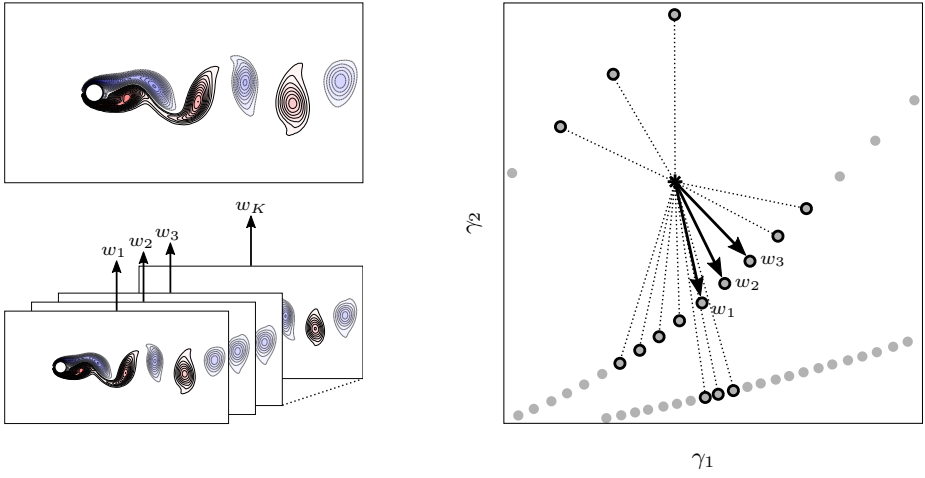
\caption[Schematic of the encoding routine]
{Schematic of the encoding step from a new snapshot $\bm{u}$.
Left: find the $K$ nearest neighbours $\bm{u}^{m_k}$, $k=1,\ldots,K$ 
      in the physical domain and compute reconstruction weights $w_1$ to $w_K$. 
Right: apply weights to compute the  corresponding feature vector
      $\bm{\gamma} = \sum_{k=1}^K w_k \bm{\gamma}^{m_k}$.}
	\label{Fig:encodingStep}
	\centering
\end{figure}
Finally,  the LLE-based encoding and decoding  for new snapshots or new feature vectors is explained.
Figure \ref{Fig:encodingStep} depicts this encoding step.
In this step, the new flow snapshot $\bm{u}$ 
is mapped into the precomputed feature-space embedding. 
At first, the $K$ nearest neighbours  $\bm{u}^{m_k}$, $k=1,\ldots,K$ are identified.
The snapshot is reconstructed by the $K$ neighbouring snapshots
as expansion $\bm{u} = \sum_{k=1}^K w_k \bm{u}^{m_k}$ using the best weights $w_i$
following the very idea of LLE.
Secondly, the new feature vector $\bm{\gamma} = \sum_{k=1}^K w_k \bm{\gamma}^{m_k}$ 
is constructed from the feature vectors of these neighbours using the same weights.
This encoding procedure marks a mapping 
from an arbitrary velocity field to a feature vector.
Note that this encoding is much more similar to clustering than to POD.
Clustering identifies the nearest centroid in the data,
i.e.\ the encoder can never `leave' the training data.
In contrast, the representation by a POD expansion
may be arbitrarily far away from the training data,
as long as it stays on the corresponding subspace.

\begin{figure}
	\centering
	\def\svgwidth{0.9\linewidth}
	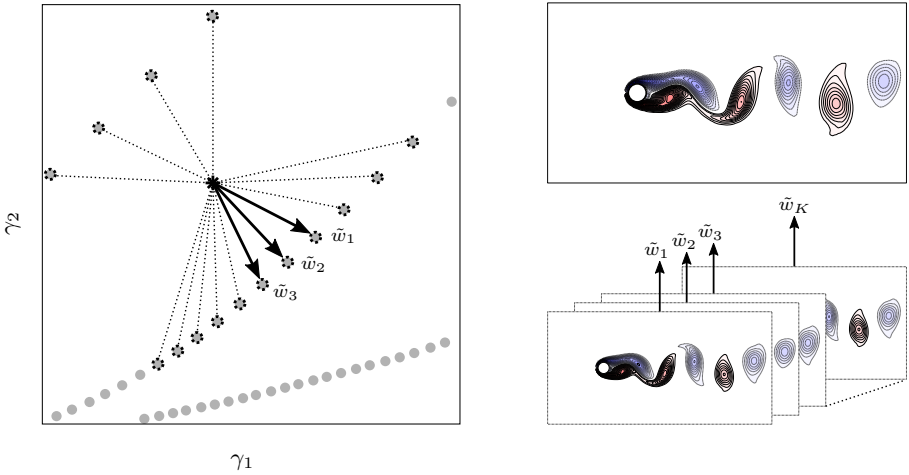
\caption[Schematic of the decoding routine]
{Schematic of the decoding step from a new feature vector $\bm{\gamma}$. 
 Left: find the $K$ nearest neighbours $\bm{\gamma}^{m_k}$, $k=1,\ldots,K$  in the feature space
       and compute the reconstruction weights $\tilde{w}_1$ to $\tilde{w}_K$. 
 Right: apply weights to original space for constructed snapshot
$\bm{u} = \sum_{k=1}^K \tilde{w}_k \bm{u}^{m_k}$.}
	\label{Fig:decodingStep}
	\centering
\end{figure}
Decoding works in the opposite direction as illustrated in figure \ref{Fig:decodingStep}. 
Given a feature vector $\bm{\gamma}$, the $K$ nearest neighbours and corresponding reconstruction weights $w_k$ are identified.
The same weights are employed for the reconstruction of the corresponding snapshot.
Note that neighbours and the reconstruction weights are not necessarily  same as in the encoding step.
The pair of encoder and decoder define an LLE-based autoencoder.
In the following section \ref{Sec:Discussion}, 
the LLE-based autoencoder is compared to POD and cluster representations.

\section{Discussion}
\label{Sec:Discussion}
In this section, 
the accuracy of the LLE-, POD- and cluster-based autoencoders
are investigated and compared 
for new transients not included on the calibration.
First (\S~\ref{Sec:ErrorRe100}), 
the LLE-based autoencoder is investigated for a transient 
with new initial conditions at the same Reynolds number.
The LLE-based representation is compared with
clustering in \S~\ref{Sec:ErrorClustering}
and a POD expansion in \S~\ref{Sec:ErrorPOD}.
Finally, the low-dimensional representations of LLE, clustering and POD 
are assessed for transient data at other Reynolds numbers (\S~\ref{Toc:ErrorRe80And120}).

\subsection{Reconstruction error for new initial conditions}
\label{Sec:ErrorRe100}
Starting point is a simulation at design Reynolds number $Re=100$
with an initial condition close to the steady solution \eqref{Eqn:InitialCondition}.
The chosen phase $\phi=12.5^\circ$  is maximally different 
from the phases $\phi = i \> 22.5^\circ$, $i=1,\ldots,16$
employed for the training data.
Thus, the corresponding snapshots represent validation data
for the out-of-sample error.
The instantaneous reconstruction error 
for out-of-sample data $E_{out}$ 
is based on the Hilbert space norm,
\begin{equation}
E_{out}=\left\Vert\bm{u}^m-\hat{\bm{u}}^m\right\Vert_{\Omega}^2,
\label{Eqn:OutOfSampleError}
\end{equation}
where $\hat{\bm{u}}^m$ denotes the low-dimensional representation of the $m$-th snapshot. 

Three different autoencoders are used for the low-dimensional representations.
LLE-based representations follow the algorithm of section \ref{Sec:Results:Autencoder}.
The cluster-based reconstruction $\hat{\bm{u}}^m$ 
is equal to the centroid that is closest to the snapshot $\bm{u}^m$.
POD-based approximations are centered at the steady solutions and are obtained by Galerkin projection.
The LLE-based autoencoder is based on a two-dimensional manifold.
For POD and clustering, we take 10 modes and centroids, respectively.
POD- and cluster-based autoencoders employ the same amount of flow states 
and may hence be considered comparable.
However, cluster-based approximation makes uneconomical use of the centroids
by not allowing for interpolations.
The two-dimensional LLE and ten-dimensional POD approximation are far from comparable.
On the one hand, compression to two feature coordinates is much higher 
for LLE as compared to ten POD mode amplitudes.
This would lead to significant errors for high-dimensional turbulent dynamics.
On the other hand, LLE uses all original snapshots 
and thus more than 112 times more flow states than the POD expansion.

\begin{figure}
	\centering
	\input{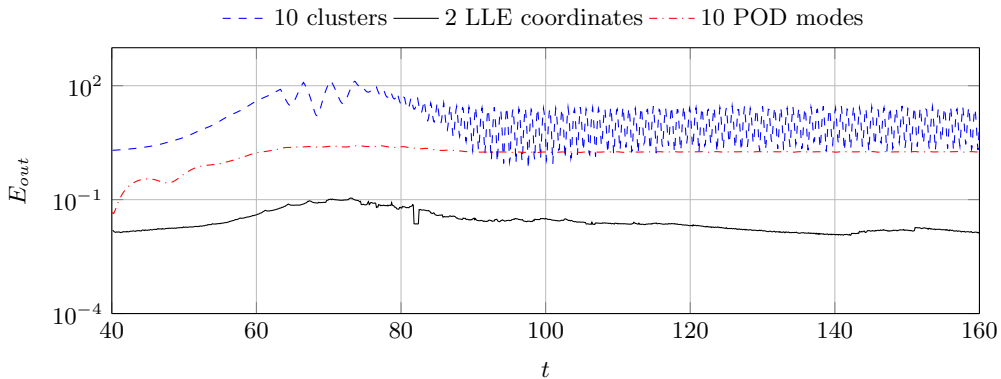}
	\caption{Out-of-sample error $E_{out}$ for a new simulation trajectory at $Re=100$.
The solid line corresponds to LLE representations.
The red dash-dotted curve and blue dashed curve refer to approximations
with 10 centroids and 10 POD modes, respectively.} 
	\label{Fig:ErrorRe100}
	\centering
\end{figure}

Figure \ref{Fig:ErrorRe100} displays the reconstruction error 
of the three methods for the new simulation data.
All three methods have the largest reconstructing error
in the transient phase between $t=60$ and $t=80$.
LLE significantly outperforms both POD and clustering by up to three orders of magnitudes, 
highlighting the two-dimensional manifold of the Navier-Stokes dynamics
and a niche application of LLE.
As expected clustering performs worst lacking any intrinsic interpolation.

\subsection{LLE and clustering}
\label{Sec:ErrorClustering}
\begin{figure}
	\centering
	\def\svgwidth{0.45\linewidth}
	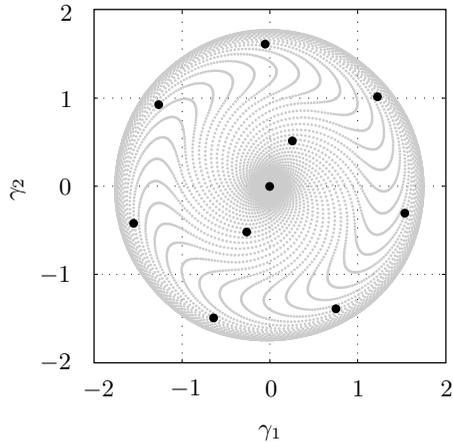
\caption{Cluster centroids localized in the LLE-based feature space.
One centroid represents the steady state solution, 
two resolve the opposite transient phases and 
the remaining eight centroids are close to the limit cycle.}
	\label{Fig:LLECluster}
\end{figure}
The  out-of-sample error of clustering  can easily be explained in the framework of LLE.
Figure \ref{Fig:LLECluster} shows the projection of the ten cluster centroids into the embedding. 
One centroid is located at the origin representing the steady solution.
Only two centroids represent the early stages of vortex shedding,
i.e.\ half a period of vortex shedding at different fluctuation levels
is represented by a single centroid.
And eight centroids resolve post-transient vortex shedding,
corresponding to large phase bins of $45^\circ$ intervals.
Thus, the large representation error of the reconstruction error
around 50\% of the fluctuation level in figure \ref{Fig:ErrorRe100}
can be attributed to the inadequate coverage of this area by the centroids.

\subsection{LLE and POD}
\label{Sec:ErrorPOD}
The post-transient POD modes of the periodic cylinder wake
are known to closely resemble real and imaginary parts of the Fourier modes.
These modes can easily be reconstructed from LLE
without the need for time-resolving snapshots.

\begin{figure}
	\centering
	\def\svgwidth{0.8\linewidth}
	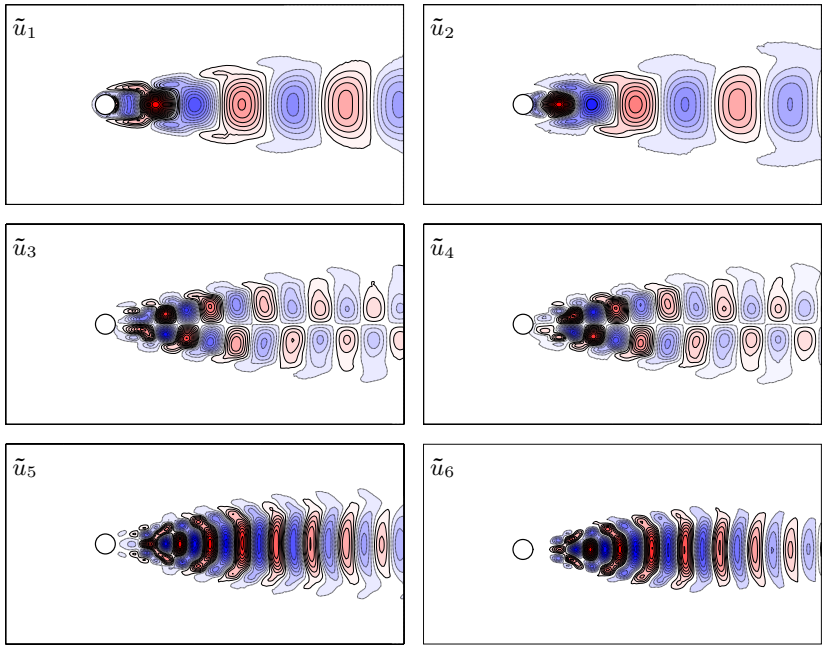
\caption{First three harmonic pairs of the post-transient cylinder wake derived from LLE and corresponding vorticity snapshots.
Note that the color scale is not equal for the modes.
Colorbars are omitted since only the qualitative structure of the modes is important.}
	\label{Fig:LLEHarmonics}
	\centering
\end{figure}
Let $r, \phi$ be 
the polar coordinates of the feature vector $\bm{\gamma}$,
i.e.\ 
$r \exp{\imath \phi} = \gamma_1 + \imath \gamma_2$.
For the Fourier mode reconstruction,
we assume a uniform rotation  $\phi = \omega t$
and hence uniform phase distribution on the limit cycle.
We only employ snapshots $\bm{u}^m$
when the corresponding feature vector $\bm{\gamma}^m$ 
with angle $\phi^m$ is on the limit cycle.
In practice, the last $300$ samples from each 16 simulations are taken.
The post-transient  $\cos$ and $\sin$ modes read
\begin{equation}
\bm{\tilde{u}}_1=\sum\limits_{m} 
\cos(\phi^m) \> (\bm{u}^m-\bm{u}_0)
\label{Eqn:CosMode}
\end{equation}
for the first phase of the first harmonic and
\begin{equation}
\bm{\tilde{u}}_2=\sum\limits_{m}
\sin(\phi^m) \> (\bm{u}^m-\bm{u}_0)
\label{Eqn:SinMode}
\end{equation}
for the second phase of the first harmonic.
The first mode  \eqref{Eqn:CosMode} 
and second mode \eqref{Eqn:SinMode} 
are most correlated with $\gamma_1$ and $\gamma_2$, respectively.
With proper normalization, both modes together resolve the first harmonics.
For accuracy reasons, 
the fluctuation around the mean flow  $\bm{u}_0$ is considered,
such that homogeneous boundary conditions are exactly fulfilled.

Similarly, higher harmonics modes can be constructed
employing $\cos ( 2 \phi^m )$, $\sin (2 \phi^m)$, etc.
Figure \ref{Fig:LLEHarmonics} displays the first three harmonic pairs derived from this approach.
Modulo phase shifts, these modes represent POD and Fourier modes.
Any other phasor $\phi$, e.g. from the lift coefficient and its derivative, 
would have performed a similar job.
LLE comes with the additional advantage 
that it also allows to construct amplitude dependent modes
by similar operations on smaller limit cycle $r=\hbox{const} \le 1.75$.
Amplitude-dependent modes can significantly improve the accuracy of empirical Galerkin models
\citep{Morzynski2006aiaa,Loiseau2019dgruyter}.

\subsection{Reconstruction error for new Reynolds numbers}
\label{Toc:ErrorRe80And120}
\begin{figure}
	\centering

	\input{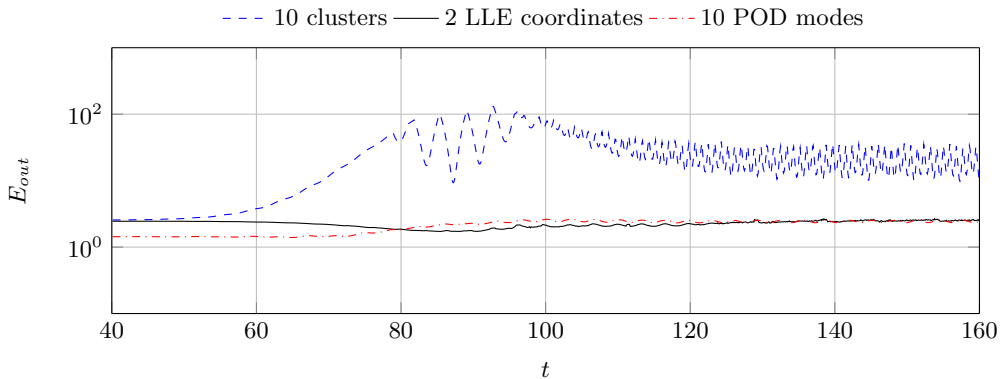}
	\caption{Same as figure \ref{Fig:ErrorRe100} 
                 but for a new simulation trajectory at $Re=80$.
LLE, cluster and POD representations are obtained from the training data at $Re=100$.
} 
	\label{Fig:ErrorRe80}
	\centering
\end{figure}
\begin{figure}
	\centering
	\input{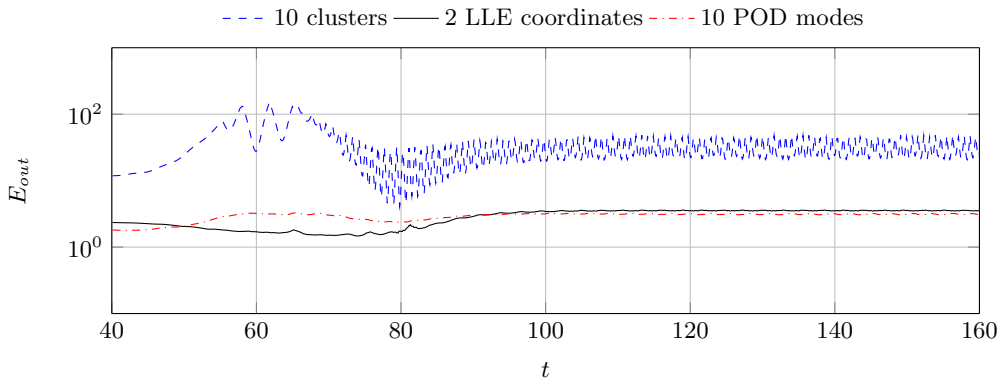}
	\caption{Same as figure \ref{Fig:ErrorRe80}
but at $Re=120$ instead of $Re=80$.}
	\label{Fig:ErrorRe120}
	\centering
\end{figure}
Finally, 
a more challenging investigation of out-of-design data is performed. 
Two validation transients with same initial condition 
but at $Re=80$ and $Re=120$ are computed.
The corresponding errors are shown in figures \ref{Fig:ErrorRe80} and \ref{Fig:ErrorRe120}.
Again,
clustering performs worst, 
suffering both from off-design flow structures
and a very coarse coverage of the manifold.
On the limit cycle, LLE and POD-based autoencoders have similar accuracy.
During the transient phase at $Re=80$,
POD performs better than LLE
because one of the POD modes resolves the mean flow variation
and can extrapolate from the training data.
The initial vortex bubble is smaller at $Re=80$ than at $Re=100$. 
At $Re=120$ this initial performance advantage of POD vanishes quickly
because LLE's good resolution of mode deformations
is more important than the smaller change of short-term averaged mean flow.

\section{Conclusions}
\label{Sec:Conclusions}

We present
arguably 
the first data-driven two-dimensional manifold representation 
of the transient oscillatory cylinder wake
using an unsupervised machine learning method.
Starting point is an ensemble of flow snapshots
from transient wake simulations.
The key enabler is the Locally Linear Embedding (LLE) encoder.
LLE has been augmented to an autoencoder with the $K$-nearest-neighbour method.
The error for new data is less than 1\%.

This accuracy strongly indicates an exactly two-dimensional Navier-Stokes dynamics
from the steady solution to the limit cycle.
A two-dimensional manifold is predicted by mean-field theory.
Yet, mean-field theory neglects the strong deformation of vortex shedding modes
from stability eigenmodes near the steady solution
to proper orthogonal decomposition (POD) modes on the limit cycle.
This mode deformation makes POD expansions 
very inefficient---requiring a 50th order expansion for a similar accuracy.

Literature presents a rich set of other methods for manifold learning \citep{Gorban2005book}.
For the same flow configuration, 
\citet{Loiseau2018jfm} present two-dimensional feature-based manifolds
from drag coefficients with similar accuracy.
Here, the engineering parameter is the choice of the feature space,
anticipating already core results of mean-field theory.
Similarly, Grassmann manifolds can be constructed from the flow data \citep{Franz2017springer,Loiseau2019dgruyter}.
Again, prior knowledge in form of the mode dependency on the shift-mode amplitude was assumed.
Look-up table approaches discretely mimic the continuous mode deformations on the manifold
by employing second-order POD mode expansions in certain parameter ranges
\citep{Lehmann2005cdc,Siegel2008jfm}.
The discrete change of vortex shedding modes is avoided
in continuous mode interpolation \citep{Morzynski2006aiaa} 
and time-dependent modes \citep{Babaee2016ptrs}.

The proposed LLE autoencoder 
is an unsupervised and purely data-driven manifold learner.
The application of LLE only assumes adequate input data covering the manifold,
but requests no advance knowledge of any structure of this manifold.
Even, the dimension of the manifold can be automatically tested
and does not have to be assumed.
The dimension can either be derived from the LLE eigenvalue problem
or can be tested from the representation residual, 
stopping the dimension increase before diminishing returns.
Thus, not only oscillatory mean-field manifolds can be approximated with LLE.
Other examples include one-dimensional homoclinic orbits,
one-dimensional heteroclinic orbits, 
two- or three-dimensional tori from quasi-periodic dynamics 
with two or three incommensurable orbits,
just to mention a few options of regular solutions.
Aerodynamic stall of pitching airfoils, 
unsteady fluid-structure interactions, and other flows
with pronounced vortices are predestined for future LLE-based representations.
\citet{Bourgeois2013jfm} report on two-dimensional manifolds
from filtered three-dimensional turbulent wake data, 
indicating application beyond regular dynamics 
and highly structured vortical flows.
Another application of LLE is the low-dimensional encoding 
of parametric dependencies of steady RANS solutions.
Thus, underlining key order parameters may be automatically distilled
from a large number of design parameters.

One cannot overemphasize the advantage of least-order dynamical models
for estimation, modeling and control design \citep{Rowley2006arfm,Noack2008jnet,Brunton2015amr}.
Any extra  state space coordinate 
acts as noise amplifier and possible direction 
where the modelled dynamics can go astray.
This is evident for estimation and dynamic modeling.
It is even more crucial for model-based control: 
The optimized feedback laws have the tendency 
to leave the domain of validity of the model,
thus predicting good model-based performance
while creating unsatisfactory results in the intended plant.
For oscillatory dynamics, the two-dimensional manifold
is not only convenient for modeling and control.
The whole manifold is also the domain of model validity
for not too aggressive control \citep{Gerhard2003aiaa}.

The current study hints at a promising avenue 
with numerous niche application of LLE
as enabler for reduced-order modeling.
LLE may  feature unique advantages over  POD, DMD, CROM and variants thereof.
The LLE-based auto-encoder can easily be upgraded to a dynamical model
by local Galerkin projection \citep{Fletcher1984book}, 
by the rich arsenal of model identification techniques
or by sparse human-interpretable models \citep{Brunton2016pnas}.
The authors actively explore the LLE avenue 
for modeling and control and for other configurations.

\section*{Acknowledgements}

This work is partically supported
by the Bernd Noack cybernetic foundation, 
by the French National Research Agency (ANR)
grants 'ACTIV\_ROAD' and 'FlowCon' (ANR-17-ASTR-0022),
and by Polish Ministry of Science and Higher Education (MNiSW) under the Grant No.: 05/54/DSPB/6492.

We appreciate valuable stimulating discussions with Markus Abel, 
Steven Brunton, Guy Cornejo-Maceda, Nan Deng, Rishabh Ishar, Eurika Kaiser, Fran\c{c}ois Lusseyran, 
Jean-Christophe Loiseau, Aditya Nair, Luc Pastur, Richard Semaan and Kunihiko (Sam) Taira.

\appendix


\begin{thebibliography}{44}
\expandafter\ifx\csname natexlab\endcsname\relax\def\natexlab#1{#1}\fi
\def\au#1{#1} \def\ed#1{#1} \def\yr#1{#1}\def\at#1{#1}\def\jt#1{\textit{#1}}
  \def\bt#1{#1}\def\bvol#1{\textbf{#1}} \def\vol#1{#1} \def\pg#1{#1}
  \def\publ#1{#1}\def\arxiv#1{#1}\def\org#1{#1}\def\st#1{\textit{#1}}

\bibitem[Abu-Mostafa {\em et~al.\/}(2012)Abu-Mostafa, Magndon-Ismail \&
  Lin]{AbuMostafa2012book}
{\sc \au{Abu-Mostafa, Y.~S.}, \au{Magndon-Ismail, M.} \& \au{Lin, H.-T.}}
  \yr{2012} {\em Learning from Data. A Short Course\/}.  \publ{AMLBook}.

\bibitem[Arthur \& Vassilvitskii(2007)]{Arthur2007proc}
{\sc \au{Arthur, D.} \& \au{Vassilvitskii, S.}} \yr{2007}  \at{k-means++: The
  advantages of careful seeding}.  \bt{In {\em Proc. of the 18th Annual
  ACM-SIAM Symposium on Discrete Algorithms\/}},  \pg{pp. 1027--1035}.
  \publ{Philadelphia, PA, USA: Society for Industrial and Applied Mathematics}.

\bibitem[Aubry {\em et~al.\/}(1988)Aubry, Holmes, Lumley \&
  Stone]{Aubry1988jfm}
{\sc \au{Aubry, N.}, \au{Holmes, P.}, \au{Lumley, J.~L.} \& \au{Stone, E.}}
  \yr{1988}  \at{The dynamics of coherent structures in the wall region of a
  turbulent boundary layer}.  \jt{J.\ Fluid Mech.}  \bvol{192},  \pg{115--173}.

\bibitem[Babaee \& Sapsis(2016)]{Babaee2016ptrs}
{\sc \au{Babaee, H.} \& \au{Sapsis, T.~P.}} \yr{2016}  \at{A variational
  principle for the description of time-dependent modes associated with
  transient instabilities}.  \jt{Phil.\ Trans.\ Roy.\ S.\ Lond.} \bvol{472}, article \pg{20150779}.

\bibitem[Barkley \& Henderson(1996)]{Barkley1996jfm}
{\sc \au{Barkley, D.} \& \au{Henderson, R.D.}} \yr{1996}  \at{Three-dimensional
  {F}loquet stability analysis of the wake of a circular cylinder}.  \jt{J.\
  Fluid Mech.}  \bvol{322},  \pg{215--241}.

\bibitem[Bourgeois {\em et~al.\/}(2013)Bourgeois, Martinuzzi \&
  Noack]{Bourgeois2013jfm}
{\sc \au{Bourgeois, J.~A.}, \au{Martinuzzi, R.~J.} \& \au{Noack, B.~R.}}
  \yr{2013}  \at{Generalised phase average with applications to sensor-based
  flow estimation of the wall-mounted square cylinder wake}.  \jt{J.\ Fluid
  Mech.}  \bvol{736},  \pg{316--350}.

\bibitem[Brunton \& Noack(2015)]{Brunton2015amr}
{\sc \au{Brunton, S.~L.} \& \au{Noack, B.~R.}} \yr{2015}  \at{Closed-loop
  turbulence control: {P}rogress and challenges}.  \jt{Appl.\ Mech.\ Rev.}
  \bvol{67}~(5),  \pg{050801:01--48}.

\bibitem[Brunton {\em et~al.\/}(2016)Brunton, Proctor \& Kutz]{Brunton2016pnas}
{\sc \au{Brunton, S.~L.}, \au{Proctor, J.~L.} \& \au{Kutz, N.~J.}} \yr{2016}
  \at{Discovering governing equations from data by sparse identification of
  nonlinear dynamical systems}.  \jt{Proc.\ Natl.\ Acad. Sci. USA}
  \bvol{113}~(5),  \pg{3932--3937}.

\bibitem[Cox \& Cox(2000)]{Cox2000book}
{\sc \au{Cox, T.~F.} \& \au{Cox, M.~A.~A.}} \yr{2000} {\em {M}ultidimensional
  {S}caling\/}, 2nd edn.,  \st{Monographs on Statistics and Applied
  Probability},  \vol{vol.~88}.  \publ{Chapman and Hall}.

\bibitem[Deane {\em et~al.\/}(1991)Deane, Kevrekidis, Karniadakis \&
  Orszag]{Deane1991pfa}
{\sc \au{Deane, A.~E.}, \au{Kevrekidis, I.~G.}, \au{Karniadakis, G.~E.} \&
  \au{Orszag, S.~A.}} \yr{1991}  \at{Low-dimensional models for complex
  geometry flows: Application to grooved channels and circular cylinders}.
  \jt{Phys.\ Fluids A}  \bvol{3},  \pg{2337--2354}.

\bibitem[Fax\'en(1927)]{Faxen1927}
{\sc \au{Fax\'en, H.}} \yr{1927}  \at{Exakte {L}\"osungen der {O}seen\-schen
  {D}ifferential\-glei\-chungen einer z\"ahen {F}l\"ussigkeit f\"ur den {F}all
  der {T}rans\-lations\-bewegung eines {Z}ylinders}.  \jt{Nova Acta Regiae
  Societiarum Upsaliensis, Volumen extra ordinem, 1--55} .

\bibitem[Fletcher(1984)]{Fletcher1984book}
{\sc \au{Fletcher, C.~A.~J.}} \yr{1984} {\em Computational Galerkin Methods\/},
  1st edn.  \publ{New York: Springer}.

\bibitem[F\"oppl(1913)]{Foeppl1913}
{\sc \au{F\"oppl, L}} \yr{1913}  \at{Wirbelbewegung hinter einen
  {K}reiszylinder}.  \jt{Sitzb.\ d.\ k.\ bayr.\ Akad.\ d.\ Wiss.}  \bvol{1},
  \pg{1--18}.

\bibitem[Franz {\em et~al.\/}(2017)Franz, Zimmermann \&
  Goertz]{Franz2017springer}
{\sc \au{Franz, T.}, \au{Zimmermann, R.} \& \au{Goertz, S.}} \yr{2017} Adaptive
  sampling for nonlinear dimensionality reduction based on manifold learning.
  \bt{In {\em Model Reduction of Parametrized Systems\/} (ed. \ed{P.~Brenner,
  M.~Ohlberger, A.~Patera, G.~Rozza \& K.~Urban})},  \pg{pp. 225--269}.
  \publ{Germany: Springer}.

\bibitem[Gerhard {\em et~al.\/}(2003)Gerhard, Pastoor, King, Noack, Dillmann,
  Morzy\'nski \& Tadmor]{Gerhard2003aiaa}
{\sc \au{Gerhard, J.}, \au{Pastoor, M.}, \au{King, R.}, \au{Noack, B.~R.},
  \au{Dillmann, A.}, \au{Morzy\'nski, M.} \& \au{Tadmor, G.}} \yr{2003}
  Model-based control of vortex shedding using low-dimensional {G}alerkin
  models.  \bt{In {\em 33rd AIAA Fluids Conference and Exhibit\/}}. Orlando,
  Florida, USA,  paper 2003-4262.

\bibitem[Gorban \& Karlin(2005)]{Gorban2005book}
{\sc \au{Gorban, A.~N.} \& \au{Karlin, I.~V.}} \yr{2005} {\em Invariant
  Manifolds for Physical and Chemical Kinetics\/}. {\em Lecture Notes in
  Physics\/} Vol.\ 660.  \publ{Berlin: Springer-Verlag}.

\bibitem[Holmes {\em et~al.\/}(2012)Holmes, Lumley, Berkooz \&
  Rowley]{Holmes2012book}
{\sc \au{Holmes, P.}, \au{Lumley, J.~L.}, \au{Berkooz, G.} \& \au{Rowley,
  C.~W.}} \yr{2012} {\em Turbulence, Coherent Structures, Dynamical Systems and
  Symmetry\/}, 2nd edn.  \publ{Cambridge: Cambridge University Press}.

\bibitem[Jackson(1987)]{Jackson1987jfm}
{\sc \au{Jackson, C.~P.}} \yr{1987}  \at{A finite-element study of the onset of
  vortex shedding in flow past variously shaped bodies}.  \jt{J.\ Fluid Mech.}
  \bvol{182},  \pg{23--45}.

\bibitem[Kaiser {\em et~al.\/}(2017)Kaiser, Noack, Spohn, Cattafesta \&
  Morzy\'nski]{Kaiser2017tcfd}
{\sc \au{Kaiser, E.}, \au{Noack, B.~R.}, \au{Spohn, A.}, \au{Cattafesta, L.~N.}
  \& \au{Morzy\'nski, M.}} \yr{2017}  \at{Cluster-based control of nonlinear
  dynamics}.  \jt{Theor.\ Comput.\ Fluid Dyn.}  \bvol{31}~(5--6),
  \pg{1579--593}.

\bibitem[von K\'arm\'an(1911)]{Karman1911}
{\sc \au{von K\'arm\'an, Th.}} \yr{1911} {\em \"Uber den {M}echanismus des
  {W}iderstandes, den ein bewegter {K}\"orper in einer {F}l\"ussigkeit
  erf\"ahrt\/},  \pg{pp. 509--517}.

\bibitem[Landau(1944)]{Landau1944}
{\sc \au{Landau, L.~D.}} \yr{1944}  \at{On the problem of turbulence}.
  \jt{C.R.\ Acad.\ Sci.\ USSR}  \bvol{44},  \pg{311--314}.

\bibitem[Lehmann {\em et~al.\/}(2005)Lehmann, Luchtenburg, Noack, King,
  Morzy\'nski \& Tadmor]{Lehmann2005cdc}
{\sc \au{Lehmann, O.}, \au{Luchtenburg, M.}, \au{Noack, B.~R.}, \au{King, R.},
  \au{Morzy\'nski, M.} \& \au{Tadmor, G.}} \yr{2005} Wake stabilization using
  {POD} {G}alerkin models with interpolated modes.  \bt{In {\em 44th IEEE
  Conference on Decision and Control and European Control Conference ECC\/}}.
  Seville, Spain, 12--15 Dec.\ 2005, invited Paper MoA15.2.

\bibitem[Lin(1954)]{Lin1954proc}
{\sc \au{Lin, C.~C.}} \yr{1954}  \at{On periodically oscillating wakes in the
  {O}seen approximation}.  \bt{In {\em Studies in Math.\ and Mech.\ presented
  to R.\ von Mises\/}},  \pg{pp. 170--176}.  \publ{New York: Academic Press}.

\bibitem[Loiseau {\em et~al.\/}(2019)Loiseau, Brunton \&
  Noack]{Loiseau2019dgruyter}
{\sc \au{Loiseau, J.-C.}, \au{Brunton, S.~L.} \& \au{Noack, B.~R.}} \yr{2019}
  From the {POD}-{G}alerkin method to sparse manifold models.  \bt{In {\em
  Handbook of Model-Order Reduction. Volume 2: Applications\/} (ed.
  \ed{P.~Benner})},  \pg{pp. 1--47 (in print)}.  \publ{DGruyter}.

\bibitem[Loiseau {\em et~al.\/}(2018)Loiseau, Noack \& Brunton]{Loiseau2018jfm}
{\sc \au{Loiseau, J.-Ch.}, \au{Noack, B.~R.} \& \au{Brunton, S.~L.}} \yr{2018}
  \at{Sparse reduced-order modeling: Sensor-based dynamics to full-state
  estimation}.  \jt{J.~Fluid Mech.}  \bvol{844},  \pg{459--490}.

\bibitem[Ma \& Karniadakis(2002)]{Ma2002jfm}
{\sc \au{Ma, X.} \& \au{Karniadakis, G.~E.}} \yr{2002}  \at{A low-dimensional
  model for simulating three-dimensional cylinder flow}.  \jt{J.\ Fluid Mech.}
  \bvol{458},  \pg{181--190}.

\bibitem[Morzy\'nski {\em et~al.\/}(1999)Morzy\'nski, Afanasiev \&
  Thiele]{Morzynski1999cmame}
{\sc \au{Morzy\'nski, M.}, \au{Afanasiev, K.} \& \au{Thiele, F.}} \yr{1999}
  \at{Solution of the eigenvalue problems resulting from global non-parallel
  flow stability analysis}.  \jt{Comput.\ Meth.\ Appl.\ Mech.\ Enrgrg.}
  \bvol{169},  \pg{161--176}.

\bibitem[Morzy\'nski {\em et~al.\/}(2006)Morzy\'nski, Stankiewicz, Noack,
  Thiele \& Tadmor]{Morzynski2006aiaa}
{\sc \au{Morzy\'nski, M.}, \au{Stankiewicz, W.}, \au{Noack, B.~R.}, \au{Thiele,
  F.} \& \au{Tadmor, G.}} \yr{2006} Generalized mean-field model for flow
  control using continuous mode interpolation.  \bt{In {\em 3rd AIAA Flow
  Control Conference\/}}. Invited AIAA-Paper 2006-3488.

\bibitem[Ng(2011)]{Ng2011proc}
{\sc \au{Ng, A.}} \yr{2011}  \at{Sparse autoencoder}.  \jt{CS294A Lecture
  notes}  \bvol{72}~(2011),  \pg{1--19}.

\bibitem[Noack {\em et~al.\/}(2003)Noack, Afanasiev, Morzy\'nski, Tadmor \&
  Thiele]{Noack2003jfm}
{\sc \au{Noack, B.~R.}, \au{Afanasiev, K.}, \au{Morzy\'nski, M.}, \au{Tadmor,
  G.} \& \au{Thiele, F.}} \yr{2003}  \at{A hierarchy of low-dimensional models
  for the transient and post-transient cylinder wake}.  \jt{J.\ Fluid Mech.}
  \bvol{497},  \pg{335--363}.

\bibitem[Noack {\em et~al.\/}(2008)Noack, Schlegel, Ahlborn, Mutschke,
  Morzy\'nski, Comte \& Tadmor]{Noack2008jnet}
{\sc \au{Noack, B.~R.}, \au{Schlegel, M.}, \au{Ahlborn, B.}, \au{Mutschke, G.},
  \au{Morzy\'nski, M.}, \au{Comte, P.} \& \au{Tadmor, G.}} \yr{2008}  \at{A
  finite-time thermodynamics of unsteady fluid flows}.  \jt{J.\ Non-Equilibr.\
  Thermodyn.}  \bvol{33},  \pg{103--148}.

\bibitem[Noack {\em et~al.\/}(2016)Noack, Stankiewicz, Morzy\'nski \&
  Schmid]{Noack2016jfm}
{\sc \au{Noack, B.~R.}, \au{Stankiewicz, W.}, \au{Morzy\'nski, M.} \&
  \au{Schmid, P.~J.}} \yr{2016}  \at{Recursive dynamic mode decomposition of
  transient and post-transient wake flows}.  \jt{J.~Fluid Mech.}  \bvol{809},
  \pg{843--872}.

\bibitem[Protas(2004)]{Protas2004pf}
{\sc \au{Protas, B.}} \yr{2004}  \at{Linear feedback stabilization of laminar
  vortex shedding based on a point vortex model}.  \jt{Phys.\ Fluids}
  \bvol{16}~(12),  \pg{4473--4488}.

\bibitem[Roweis \& Lawrence(2000)]{Roweis2000s}
{\sc \au{Roweis, S.} \& \au{Lawrence, S.}} \yr{2000}  \at{Nonlinear
  dimensionality reduction by locally linear embedding}.  \jt{Science}
  \bvol{290}~(5500),  \pg{2323--2326}.

\bibitem[Rowley {\em et~al.\/}(2009)Rowley, Mezi\'c, Bagheri, Schlatter \&
  Henningson]{Rowley2009jfm}
{\sc \au{Rowley, C.~W.}, \au{Mezi\'c, I.}, \au{Bagheri, S.}, \au{Schlatter, P.}
  \& \au{Henningson, D.S.}} \yr{2009}  \at{Spectral analysis of nonlinear
  flows}.  \jt{J.\ Fluid Mech.}  \bvol{645},  \pg{115--127}.

\bibitem[Rowley \& Williams(2006)]{Rowley2006arfm}
{\sc \au{Rowley, C.~W.} \& \au{Williams, D.~R.}} \yr{2006}  \at{Dynamics and
  control of high-{R}eynolds number flows over open cavities}.  \jt{Ann.\ Rev.\
  Fluid Mech.}  \bvol{38},  \pg{251--276}.

\bibitem[Schmid(2010)]{Schmid2010jfm}
{\sc \au{Schmid, P.~J.}} \yr{2010}  \at{Dynamic mode decomposition for
  numerical and experimental data}.  \jt{J.\ Fluid. Mech}  \bvol{656},
  \pg{5--28}.

\bibitem[Schumm {\em et~al.\/}(1994)Schumm, Berger \& Monkewitz]{Schumm1994jfm}
{\sc \au{Schumm, M.}, \au{Berger, E.} \& \au{Monkewitz, P.A.}} \yr{1994}
  \at{Self-excited oscillations in the wake of two-dimensional bluff bodies and
  their control}.  \jt{J.\ Fluid Mech.}  \bvol{271},  \pg{17--53}.

\bibitem[Siegel {\em et~al.\/}(2008)Siegel, Seidel, Fagley, Luchtenburg, Cohen
  \& McLaughlin]{Siegel2008jfm}
{\sc \au{Siegel, S.~G.}, \au{Seidel, J.}, \au{Fagley, C.}, \au{Luchtenburg,
  D.~M.}, \au{Cohen, K.} \& \au{McLaughlin, T.}} \yr{2008}  \at{Low dimensional
  modelling of a transient cylinder wake using double proper orthogonal
  decomposition}.  \jt{J.\ Fluid Mech.}  \bvol{610},  \pg{1--42}.

\bibitem[Stuart(1958)]{Stuart1958jfm}
{\sc \au{Stuart, J.T.}} \yr{1958}  \at{On the non-linear mechanics of
  hydrodynamic stability}.  \jt{J.\ Fluid Mech.}  \bvol{4},  \pg{1--21}.

\bibitem[Timme(1959)]{Timme1959rep}
{\sc \au{Timme, A.}} \yr{1959}  \bt{\"uber die {E}igenschaften von
  {W}irbelstra{\ss}en}. {\em Tech. Rep.\/}~77.  \org{DVL}, K\"oln.

\bibitem[Zebib(1987)]{Zebib1987jem}
{\sc \au{Zebib, A.}} \yr{1987}  \at{Stability of viscous flow past a circular
  cylinder}.  \jt{J.\ Engr.\ Math.}  \bvol{21},  \pg{155--165}.

\bibitem[Zhang {\em et~al.\/}(1995)Zhang, Fey, Noack, K\"onig \&
  Eckelmann]{Zhang1995pf}
{\sc \au{Zhang, H.-Q.}, \au{Fey, U.}, \au{Noack, B.~R.}, \au{K\"onig, M.} \&
  \au{Eckelmann, H.}} \yr{1995}  \at{On the transition of the cylinder wake}.
  \jt{Phys.\ Fluids}  \bvol{7}~(4),  \pg{779--795}.

\bibitem[Zielinska \& Wesfreid(1995)]{Zielinska1995pf}
{\sc \au{Zielinska, B.J.A.} \& \au{Wesfreid, J.E.}} \yr{1995}  \at{On the
  spatial structure of global modes in wake flow}.  \jt{Phys.\ Fluids}
  \bvol{7}~(6),  \pg{1418--1424}.

\end{thebibliography}

\end{document}